\documentclass[twocolumn,showpacs,prb,aps]{revtex4}
\usepackage{amsmath}
\usepackage{graphicx}

\begin{document}

\title{All-angle zero reflection at metamaterial surfaces}
\author{Xin Li,$^1$ Zixian Liang,$^2$ Xiaohan Liu,$^{1,3}$ Xunya Jiang,$^2$
and Jian Zi$^{1,3}$\footnote{%
Electronic mail: jzi@fudan.edu.cn}}
\affiliation{$^1$Department of Physics and Surface Physics Laboratory, Fudan University,
Shanghai 200433, People's Republic of China\\
$^2$Shanghai Institute of Microsystem and Information Technology, CAS,
Shanghai 200050, People's Republic of China\\
$^3$Laboratory of Advanced Materials, Fudan University, Shanghai 200433,
People's Republic of China}
\date{\today}

\begin{abstract}
The authors study theoretically reflection on the surface of a metamaterial
with a hyperbolic dispersion. It is found that reflection is strongly
dependent on how the surface is terminated with respect to the asymptote of
the hyperbolic dispersion. For a surface terminated normally to the
asymptote, zero reflection occurs for all incident angles. It is exemplified
by a metamaterial made of a periodic metal-dielectric layered structure with
its surface properly cut through numerical simulations.
\end{abstract}

\maketitle

Metamaterials are artificially designed composites consisting of periodic
subwavelength structures. The optical response of metamaterials originates
from their structures instead from their compositions, leading to many
unusual optical properties that do not occur in nature. For instance,
metamaterials with a negative refractive index can produce negative
refraction\cite{she:01,par:03,hou:03} and superlensing.\cite%
{pen:00,grb:04,fan:05} In contrast to conventional materials, the energy
transport through negative-refractive-index metamaterials is in a direction
opposite to the phase direction, giving rise to reversed Doppler effects and
inverted Cherenkov cone.\cite{ves:68,pen:06a} Metamaterials can also be used
to construct invisible cloak.\cite{pen:06,sch:06}

Reflection is a wave phenomenon occurring for waves impinging upon a
surface. Our common understanding is that reflection is inevitable although
it can be eliminated at some special incident angles, e.g., Brewster's
angles. In this Letter, we show theoretically that all-angle zero reflection
can occur at the surface of a metamaterial with a hyperbolic dispersion. It
is exemplified by a metamaterial made of a periodic metal-dielectric layered
structure through numerical simulations.

\begin{figure}[htbp]
\centerline{\includegraphics[angle=0,width=7cm]{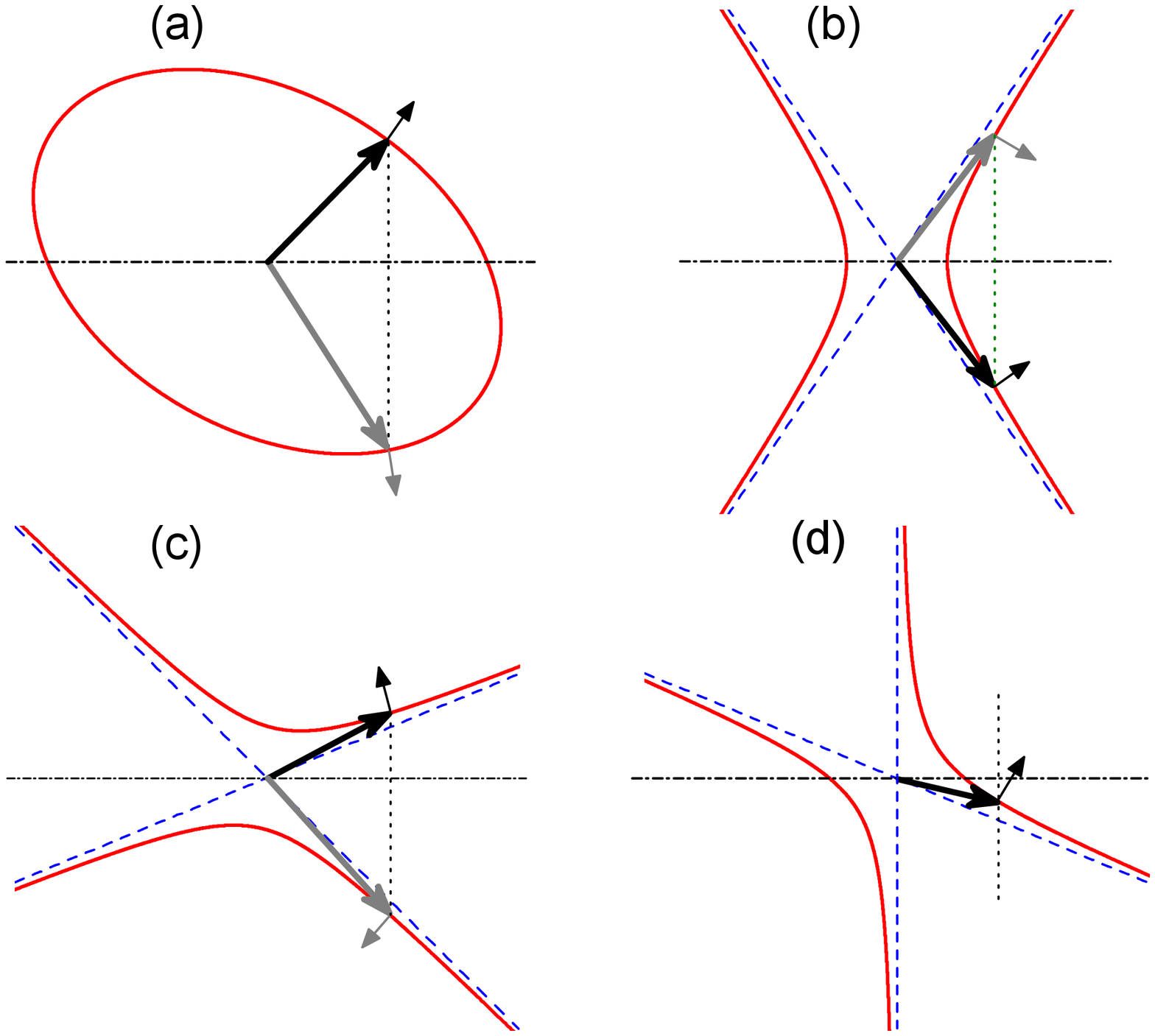}}
\caption{(Color online) Equi-frequency surface analysis of
reflection on surfaces of metamaterials with (a) an elliptic and
(b-d) a hyperbolic dispersion. Surfaces (dash-dotted lines) lie in
the horizontal plane. For metamaterials with the hyperbolic
dispersion, surfaces are cut along the principal axis (b),
perpendicularly to the asymptote (dashed lines) of the hyperbolic
dispersion (d), and obliquely to both the principal axis and
asymptote (c). Black (grey) thick arrows denote the incident
(reflected) wave vector. Thin arrows indicate the direction of the
group velocity. Dotted lines illustrate the conservation of the
in-plane wave vectors. Note that metamaterials occupy the half-space
below the horizontal line and waves are incident from the
metamaterial side. } \label{fig1}
\end{figure}

For any reflection on a surface, the spatial and time variation of
all fields must be the same at the surface. As a result, the
in-plane wave vector of an incident wave should be equal to that of
the reflected one, independent of the nature of the boundary
conditions.\cite{jac:99} For homogeneous media, this leads to the
fact that the incident angle is equal to the reflected angle. For
anisotropic media, however, incident and reflected angles may not be
the same. For either isotropic media or anisotropic media with an
elliptic dispersion, reflection always exists due to the fact that
the conservation of in-plane wave vectors of incident and reflected
waves can be always satisfied, regardless of the cutting directions
of surfaces, as shown schematically in Fig. \ref{fig1}. For an
anisotropic medium with a hyperbolic dispersion, reflection is
sensitive to the surface termination direction, namely, the surface
orientation with respect to the asymptote of the hyperbolic
dispersion. For surfaces terminated along the principal axes,
reflection always exists with equal incident and reflected angles.
For surfaces oriented obliquely to both the asymptotes and principal
axes, the reflected wave vector still exists with the reflected
angle different from the incident angle. If the surface is cut
perpendicularly to the asymptote, however, we cannot find reflected
wave vector for any given incident wave vector. In this situation,
zero reflection is expected for all incident angles as to be shown
later. It should be noted that the energy flow direction is the same
as that of the group velocity, defined by $\mathbf{v}_{g}=\nabla
\omega (\mathbf{k})$, where $k$ is the wave vector and $\omega
(\mathbf{k})$ is the dispersion relation. From its definition, the
group velocity direction is perpendicular
to the equi-frequency surface and points to the direction along which $%
\omega (\mathbf{k})$ is increasing.

A qualitative analysis of all-angle zero reflection was given
hereinbefore. In the following, we would like to give a rigorous
proof. We consider an anisotropic metamaterial whose permittivity
tensor and permeability tensor are both diagonalizable.\cite{smi:03}
To simply the proceeding analysis, we assume that the metamaterial
is nonmagnetic, namely, the diagonal elements of the permeability
tensor are all equal to 1. Without loss of generality, we assume in
our analysis a plane wave with the magnetic field polarized
along the $y$ direction with the form $\mathbf{H}=H_{0}\widehat{\mathbf{y}}%
\exp \left[ i(k_{x}x+k_{z}z)-\omega t\right] $, where $\widehat{\mathbf{y}}$
is the unit vector along the $y$ direction. If we choose a Cartesian
coordinate system $x$-$z$ with its axes along the principal axes of the
metamaterial, this plane wave satisfies the following dispersion relation%
\begin{equation}
\frac{k_{x}^{2}}{\varepsilon _{z}}+\frac{k_{z}^{2}}{\varepsilon _{x}}=\frac{%
\omega ^{2}}{c^{2}},  \label{dis}
\end{equation}%
where $\varepsilon _{x}$ and $\varepsilon _{z}$ are the diagonal elements of
the permittivity tensor. For $\varepsilon _{x}$ and $\varepsilon _{z}$ with
opposite signs, the corresponding dispersion is hyperbolic. If we choose a
new Cartesian coordinate system $x^{\prime }$-$z^{\prime }$ with the same
origin, the permittivity tensor is no longer diagonal and is transformed to%
\begin{equation}
\left[
\begin{array}{lr}
\varepsilon _{x^{\prime }x^{\prime }} & \varepsilon _{x^{\prime }z^{\prime }}
\\
\varepsilon _{z^{\prime }x^{\prime }} & \varepsilon _{z^{\prime }z^{\prime }}%
\end{array}%
\right] =\left[
\begin{array}{lr}
c^{2}\varepsilon _{x}+s^{2}\varepsilon _{z} & cs(\varepsilon
_{x}-\varepsilon _{z}) \\
cs(\varepsilon _{x}-\varepsilon _{z}) & s^{2}\varepsilon
_{x}+c^{2}\varepsilon _{z}%
\end{array}%
\right] .
\end{equation}%
Here, the parameters $c$ and $s$ are given by
\begin{equation}
c=\cos ^{2}\theta ,\text{ \ }s=\sin ^{2}\theta ,
\end{equation}
where $\theta $ is the angle between the $x^{\prime }$ and $x$ axes. The
dispersion in the new coordinate system becomes accordingly%
\begin{equation}
\frac{\varepsilon _{x^{\prime }x^{\prime }}^{2}k_{x^{\prime
}}^{2}+\varepsilon _{z^{\prime }z^{\prime }}^{2}k_{z^{\prime
}}^{2}+2\varepsilon _{x^{\prime }z^{\prime }}k_{x^{\prime }}k_{z^{\prime }}}{%
\varepsilon _{x^{\prime }x^{\prime }}\varepsilon _{z^{\prime }z^{\prime
}}-\varepsilon _{x^{\prime }z^{\prime }}^{2}}=\frac{\omega ^{2}}{c^{2}}.
\label{dis2}
\end{equation}

For a metamaterial with a hyperbolic dispersion, reflection is strongly
dependent on the cutting direction of the surface. Without loss of
generality, we assume that the metamaterial surface is terminated normally
to the $z^{\prime }$ axis, i.e., in the $x^{\prime }y^{\prime }$ plane.
Consequently, $k_{x^{\prime }}$ is the in-plane wave vector of an plane wave
and $k_{z^{\prime }}$ is the perpendicular component. For a given in-plane
wave vector $k_{x^{\prime }}$ of an incident plane wave, we can always find
a real solution of $k_{z^{\prime }}$ for the reflected wave from the
conservation of the in-plane wave vector, if the $x^{\prime }$ axis is not
perpendicular to the asymptote of the hyperbolic dispersion. For the $%
x^{\prime }$ axis just perpendicular to the asymptote, no real solution of $%
k_{z^{\prime }}$ for the reflected wave can be found. In this case, $%
k_{z^{\prime }}$ should be a complex number possessing both a real and an
imaginary part. Zero reflection is expected as can be confirmed by
calculating the normal component of the reflected Poynting vector with
respect to the surface, defined in cgs units by%
\begin{equation}
S_{r\perp }=\frac{4\pi }{c}\mathrm{Re}\left( \mathbf{E}_{r}\times \mathbf{H}%
_{r}^{\ast }\right) _{\perp }=\frac{4\pi }{c}\mathrm{Re}\left( E_{rx^{\prime
}}H_{ry^{\prime }}^{\ast }\right) ,
\end{equation}%
where $E_{r}$ and $H_{r}$ are the reflected electric and magnetic fields
with their components related each other by
\begin{equation}
E_{rx^{\prime }}=-\frac{c}{\omega }\frac{\varepsilon _{z^{\prime }z^{\prime
}}k_{z^{\prime }}+\varepsilon _{x^{\prime }z^{\prime }}k_{x^{\prime }}}{%
\varepsilon _{x^{\prime }x^{\prime }}\varepsilon _{z^{\prime }z^{\prime
}}-\varepsilon _{x^{\prime }z^{\prime }}^{2}}H_{ry^{\prime }}.
\end{equation}%
Suppose a complex $k_{z^{\prime }}$ for the reflected wave and substitute it
into Eq. (\ref{dis2}) we can obtain
\begin{equation}
\varepsilon _{z^{\prime }z^{\prime }}\mathrm{Re}(k_{z^{\prime
}})+\varepsilon _{x^{\prime }z^{\prime }}k_{x^{\prime }}=0.
\end{equation}%
This immediately leads to the fact that $E_{rx^{\prime }}$ does not possess
a real part, leading to $S_{r\perp }=0$. We can thus conclude that the
reflected wave does not carry any energy along the surface normal, implying
zero reflection.

One feasible realization of a metamaterial with a hyperbolic dispersion is
to adopt a periodic metal-dielectric layered structure.\cite%
{ram:03,bel:06,sal:06,liu:07} In the long wavelength limit (the period is
much smaller than the operating wavelength), this periodic metal-dielectric
layered structure can be viewed as a effective  anisotropic metamaterial
with the permittivity tensor given by%
\begin{equation}
\left(
\begin{array}{ccc}
\varepsilon _{x} & 0 & 0 \\
0 & \varepsilon _{x} & 0 \\
0 & 0 & \varepsilon _{z}%
\end{array}%
\right) .
\end{equation}%
For $p$-polarized waves, $\varepsilon _{x}$ and $\varepsilon _{z}$
are related to the parameters of the constituents
by\cite{ram:03,ber:78}
\begin{subequations}
\begin{eqnarray}
\varepsilon _{x} &=&\frac{\varepsilon _{1}d_{1}+\varepsilon _{2}d_{2}}{d},%
\text{ } \\
\varepsilon _{z} &=&\frac{\varepsilon _{1}\varepsilon _{2}d}{\varepsilon
_{1}d_{2}+\varepsilon _{2}d_{1}},
\end{eqnarray}%
\end{subequations}
where $\varepsilon _{1,2}$ is the dielectric constant of the constituents, $%
d_{1,2}$ is the thickness, and $d=d_{1}+d_{2}$ is the period. If we adopt a
Drude model\cite{ash:76} to describe the dielectric constant of the metal
constituent, namely, $\varepsilon (\omega )=1-\omega _{p}^{2}/\omega ^{2},$
where $\omega _{p}$ is the plasma frequency of the metal, it can be easily
shown that $\varepsilon _{x}$ and $\varepsilon _{z}$ can have opposite signs
at certain frequencies for a proper choice of the thickness parameters. This
leads to a hyperbolic dispersion for the periodic metal-dielectric layered
structure.

\begin{figure}[tbp]
\centerline{\includegraphics[angle=0,width=7cm]{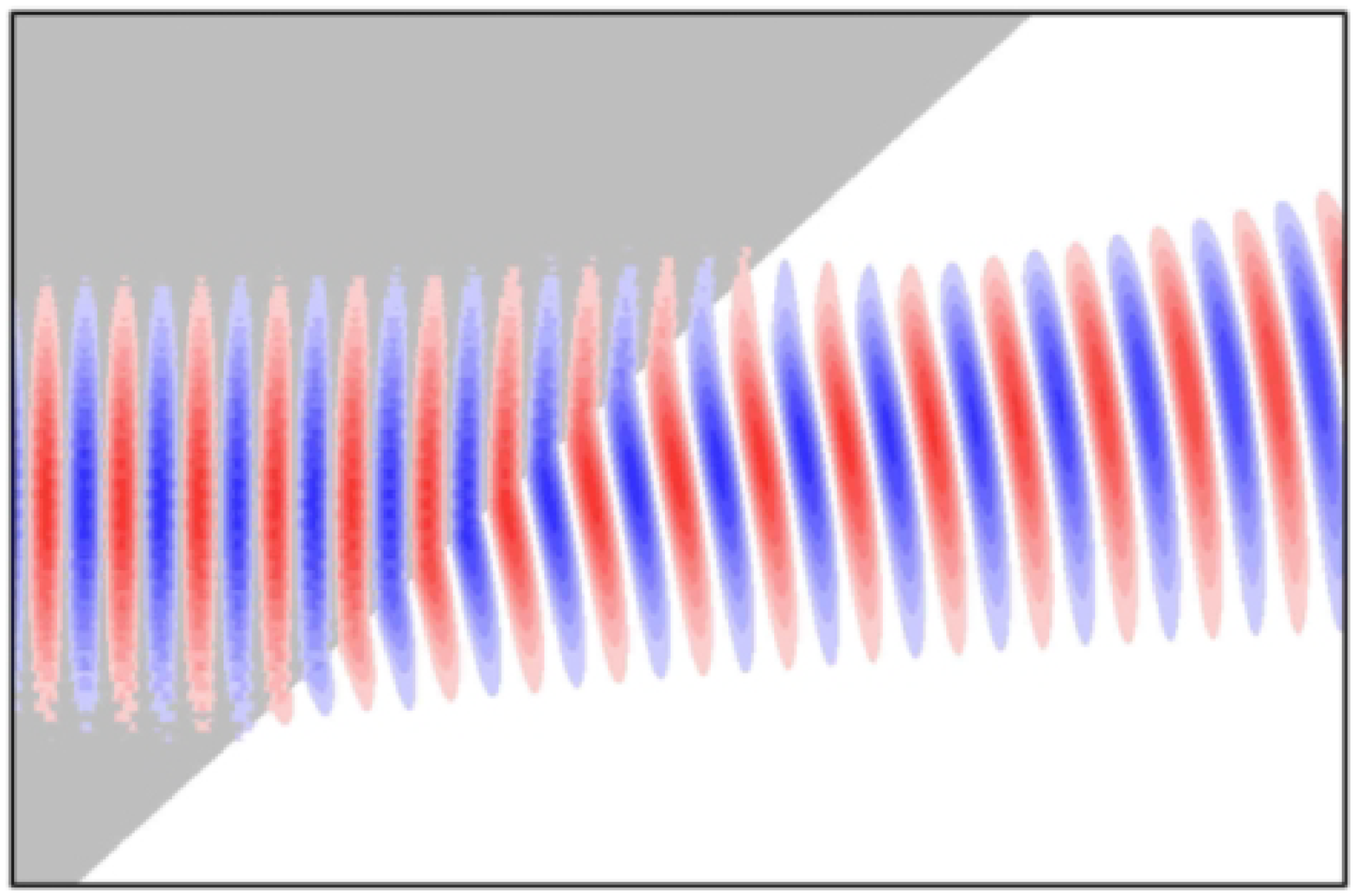}}
\caption{(Color online) FDTD simulations of the magnetic field
distributions for a $p$-polarized Gaussian beam launched from a
periodic metal-dielectric layered structure upon an interface
between the structure (grey area) and a dielectric medium (white
area) with a dielectric constant of 2.5.} \label{fig2}
\end{figure}

To illustrate all-angle zero reflection on the surface of a periodic
metal-dielectric layered structure, we carry out finite-difference
time-domain (FDTD) simulations\cite{taf:95} with perfectly matched layer
boundary conditions,\cite{ber:94} shown in Fig. \ref{fig2}. Without loss of
generality, the periodic metal-dielectric layered structure is assumed to
made from Ag and a dielectric with a dielectric constant of 2.231. The
thickness of the Ag layer is 10 nm and that of the dielectric layer is 50
nm. In our simulations, a $p$-polarized Gaussian beam is used. Its
wavelength is 756 nm, which is much larger than the period (60 nm) of the
periodic metal-dielectric layered structure, justifying our effective medium
approximation. An experimental value\cite{joh:72} of the refractive index $%
n=0.03+i5.242$ at 756 nm for Ag is used. These parameters result in Re$%
(\varepsilon _{x})=2.72$ and Re$(\varepsilon _{z})=-2.72$, leading to a
rectangularly hyperbolic dispersion for the periodic metal-dielectric
layered structure.

The terminated surface of the periodic metal-dielectric layered
structure is perpendicular to one of the asymptote of the hyperbolic
dispersion, namely, in an angle of 45$^{\circ }$ with respect to the
periodic direction. The incident Gaussian beam forms an angle of
45$^{\circ }$ with respect to the surface normal and is
perpendicular to the periodic direction. It is obvious from the FDTD
simulations that no reflection occurs at the surface. It should be
noted that zero reflection does not depend on the incident angle.
Simulations for other incident angles are also conducted and zero
reflection is always found, manifesting all-angle zero reflection.

In conclusion, the reflection on the surface of a metamaterial with a
hyperbolic dispersion is studied theoretically. For the metamaterial surface
terminated obliquely to the asymptote of the hyperbolic dispersion,
reflection is inevitable. However, all-angle zero reflection can occur if
the surface is cut perpendicularly to the asymptote. We show this by a
rigorous proof that for any incident angle the surface normal component of
the Poynting vector of the reflected wave does not possess a real part,
implying zero energy flow along the surface normal. Numerical simulations of
a periodic metal-dielectric layered structure that is cut properly affirm
unambiguously our theoretical prediction of all-angle zero reflection.
Metamaterials with all-angle zero reflection at their surfaces could be
exploited in many applications to eliminate undesirable reflection.

This work was supported by the 973 Program (grant nos. 2007CB613200 and
2006CB921700). The research of X.H.L and J.Z is further supported by NSFC
and Shanghai Science and Technology Commission is also acknowledged.


\begin{thebibliography}{99}
\bibitem{she:01} R. A. Shelby, D. R. Smith, and S. Schultz, Science \textbf{%
292}, 77 (2001).

\bibitem{par:03} C. G. Parazzoli, R. B. Greegor, K. Li, B. E. C. Koltenbah,
and M. Tanielian, Phys. Rev. Lett. \textbf{90}, 107401 (2003).

\bibitem{hou:03} A. A. Houck, J. B. Brock, and I. L. Chuang, Phys. Rev.
Lett. \textbf{90}, 137401 (2003).

\bibitem{pen:00} J. B. Pendry, Phys. Rev. Lett. \textbf{85}, 3966 (2000).

\bibitem{grb:04} A. Grbic, and G. V. Eleftheriades, Phys. Rev. Lett. \textbf{%
92}, 117403 (2004).

\bibitem{fan:05} N. Fang, H. Lee, C. Sun, and X. Zhang, Science \textbf{308}%
, 534 (2005).

\bibitem{ves:68} V. G. Veselago, Sov. Phys. Usp. \textbf{10}, 509 (1968).

\bibitem{pen:06a} J. B. Pendry and D. R. Smith, Sci. Am. \textbf{295}, 60
(2006).

\bibitem{pen:06} J. B. Pendry, D. Schurig, D. R. Smith, Science \textbf{312}%
, 1780 (2006).

\bibitem{sch:06} D. Schurig, J. J. Mock, B. J. Justice, S. A. Cummer, J. B.
Pendry, A. F. Starr, D. R. Smith, Science \textbf{314}, 977 (2006).

\bibitem{jac:99} J. D. Jackson, \textit{Classical Electrodynamics, }3rd ed.
(Wiley, New York, 1999), p. 302.

\bibitem{smi:03} D. R. Smith and D. Schurig, Phys. Rev. Lett. \textbf{90},
77405 (2003).

\bibitem{ram:03} S. A. Ramakrishna, J. B. Pendry, M. C. K. Wiltshire, and W.
J. Stewart, J. Mod. Opt. \textbf{50}, 1419 (2003).

\bibitem{bel:06} P. A. Belov and Y. Hao, Phys. Rev. B \textbf{73}, 113110
(2006).

\bibitem{sal:06} A. Salandrino and N. Engheta, Phys. Rev. B \textbf{74},
075103 (2006).

\bibitem{liu:07} Z. W. Liu, H. Lee, Y. Xiong, C. Sun, and X. Zhang, Science
\textbf{315}, 1686 (2007).

\bibitem{ber:78} D. Bergman, Phys. Lett. \textbf{43}, 377 (1978).

\bibitem{ash:76} N. W. Ashcroft and N. D. Mermin, \textit{Solid State Physics%
} (Saunders, New York, 1976), p. 156.

\bibitem{taf:95} A. Taflove, \textit{Computational Electrodynamics: The
Finite-Difference Time-Domain Method} (Artech House, Boston, 1995).

\bibitem{ber:94} J. P. Berenger, J. Comput. Phys. \textbf{114}, 185 (1994).

\bibitem{joh:72} P. B. Johnson and R. W. Christy, Phys. Rev. B \textbf{6},
4370 (1972).
\end{thebibliography}
\end{document}